\documentclass[12pt]{article}
\usepackage{latexsym}
\begin{document}

\title{{\bf EXTENDED\\ ELECTRODYNAMICS}:
\\III. Free Photons and (3+1)-Soliton-like\\ Vacuum Solutions}
\author{{\bf S.Donev}\\
Institute for Nuclear Research and Nuclear Energy,\\
Bulg.Acad.Sci., 1784 Sofia, blvd.Tzarigradsko shausee 72\\
e-mail: sdonev@inrne.acad.bg\\ BULGARIA \and \\
{\bf M.Tashkova}\\
Institute of Organic Chemistry with Center of \\ Phytochemistry,
Bulg.Acad.Sci., 1574 Sofia,\\
Acad. G.Bonchev Str., Bl. 9\\BULGARIA}

\date{}

\maketitle
patt-sol/9710003
\begin{abstract}
This paper aims to give explicitly all non-linear vacuum solutions to our
non-linear field equations [1], and to define in a coordinate free
manner the important subclass of non-linear solutions, which we call
{\it almost photon-like}. By means of a correct introduction of the local and
integral {\it intrinsic angular momentums} of these solutions, we saparate
the {\it photon-like} solutions through the requirement their integral
intrinsic angular momentums to be equal to the Planck's constant $h$.
Finally, we consider such solutions, moving radially to or from a given
center, using standard spherical coordinates.

\end{abstract}

\vskip 0.05cm
\newpage
\section{Explicit non-linear vacuum solutions}
As it was shown in [1] with every nonlinear solution $F$ of
our nonlinear equations (we use all notations from [1])
\begin{equation}
\delta F\wedge *F =0,\ \delta *F \wedge **F=0,\          
\delta *F\wedge *F-\delta F\wedge F=0.
\end{equation}
a class of $F$-adapted coordinate systems is
associated, such that $F$ and $*F$ acquire the form respectively
\[
F=\varepsilon udx\wedge dz + udx\wedge d\xi + \varepsilon pdy\wedge dz +
pdy\wedge d\xi
\]
\[
*F=-pdx\wedge dz -\varepsilon pdx\wedge d\xi + udy\wedge
dz + \varepsilon udy\wedge d\xi.
\]
Since we look for non-linear solutions of (1), we substitute these $F$ and
$*F$ in (1) and after some elementary calculations we obtain
\[
\delta F=(u_\xi-\varepsilon u_z)dx +(p_\xi-\varepsilon p_z)dy +
\varepsilon(u_x + p_y)dz +(u_x + p_y)d\xi,
\]
\[
\delta *F=-\varepsilon(p_\xi -\varepsilon p_z)dx
+\varepsilon(u_\xi-\varepsilon p_z)dy - (p_x - u_y)dz - (p_x - u_y)d\xi,
\]
\[
F_{\mu\nu}(\delta F)^\nu
dx^\nu= (*F)_{\mu\nu}(\delta *F)^\nu dx^\nu=
\]
\[
=\varepsilon\left[p(p_\xi-\varepsilon p_z)+
u(u_\xi-\varepsilon u_z)\right]dz+
\left[p(p_\xi-\varepsilon p_z)+u(u_\xi-\varepsilon u_z)\right]d\xi,
\]
\[
(\delta F)^2=(\delta *F)^2=
-(u_\xi-\varepsilon u_z)^2-(p_\xi-\varepsilon p_z)^2
\]
A simple direct calculation shows, that the equation
\[
\delta *F\wedge *F-\delta F\wedge F=0
\]
is identically
fulfilled for any such $F$ and $*F$ with arbitrary $u$ and $p$.
We obtain that our equations reduce to only 1 equation, namely
\begin{equation}
p(p_\xi-\varepsilon p_z)+u(u_\xi-\varepsilon u_z)=
\frac 12\left[(u^2+p^2)_\xi-\varepsilon(u^2+p^2)_z\right]=0.
\end{equation}
The obvious solution to this equation is
\begin{equation}
u^2+p^2=\phi^2 (x,y,\xi+\varepsilon z).
\end{equation}
The solution obtained shows that the equations impose some
limitations only on
the amplitude function $\phi$ and the phase function $\varphi$ is arbitrary
exept that it is bounded: $|\varphi |\leq 1$. The amplitude $\phi$ is a
running wave along the specially chosen coordinate $z$, which is common for
all $F$-adapted coordinate systems.Considered as a function of the spatial
coordinates, the amplitude $\phi$ is {\it arbitrary}, so it can be chosen
{\it spatially finite}. The time-evolution does not affect the initial form
of $\phi$, so it will stay the same in time.
This shows, that {\it among the
nonlinear solutions of our equations there are (3+1) soliton-like
solutions}. The spatial structure is determined by the initial condition,
while the phase function $\varphi$ can be used to define {\it internal
dynamics} of the solution.

Recalling from [1] the substitutions
\[
u=\phi.\varphi, \ p=\phi\sqrt{1-\varphi^2},
\]
and the equality $|A|=\phi$, we get
\begin{equation}
|\delta F|=|\delta *F|=
\frac{|\phi||\varphi_\xi-\varepsilon\varphi_z|}
{\sqrt{1-\varphi^2}},
\ L=\frac{|A|}{|\delta F|}=
\frac{\sqrt{1-\varphi^2}}{|\varphi_\xi-\varepsilon\varphi_z|}.
\end{equation}
For the induced pseudoorthonormal bases (1-forms and vector fields)
we find
\[
{\bf A}=\varphi dx +\sqrt{1-\varphi^2}dy,\
{\bf \varepsilon A^*}=-\sqrt{1-\varphi^2}dx+\varphi dy,\ {\bf R}=
-dz,\ {\bf S}=d\xi,
\]
\[
{\bf A}=-\varphi\frac{\partial}{\partial
x}-\sqrt{1-\varphi^2}\frac{\partial}{\partial y},\ \varepsilon{\bf A^*} =
\sqrt{1-\varphi^2}\frac{\partial}{\partial x}-\varphi\frac{\partial}
{\partial y},\
{\bf R}=\frac{\partial}{\partial z},\ {\bf S}=\frac{\partial}{\partial \xi}.
\]

Hence, the nonlinear solutions in canonical coordinates are parametrized by
one function $\phi$ of 3 parameters and one {\it bounded} function of 4
parameters. Therefore, the separation of various subclasses of nonlinear
solutions is made by imposing additional conditions on these two functions.
Further we are going to separate a subclass of solutions ,
the integral properties of which reflect well enough the well known from the
experiment integral properties and characteristics of the free photons.
These
solutions will be called {\it photon-like} and will be separated through
imposing additional requirements on $\phi$ and $\varphi$ in a
coordinate-free manner.

\section{Almost photon-like solutions}
We note first, that we have three invariant quantities at hand: $\phi$,
$\varphi$ and $L$. The amplitude function $\phi$ is to be determined by the
initial conditions, which have to be {\it finite}. So, we may impose
additional conditions on $L$ and $\varphi$. These conditions {\it have to
express some intra-consistency among the various characteristics of the
solution}. The idea, what kind of intra-consistency to use, comes from the
observation that {\it the amplitude function $\phi$ is a first integral of
the vector field ${\bf V}$},
i.e.
\[
{\bf V}(\phi)=\left(-\varepsilon \frac{\partial}{\partial z}+
\frac {\partial}{\partial \xi}\right)(\phi)=
-\varepsilon \frac{\partial}{\partial
z}\phi(x,y,\xi+\varepsilon z)
+\frac {\partial}{\partial \xi}\phi(x,y,\xi+\varepsilon z)=0.
\]
We want to extend this available consistency between ${\bf V}$ and $\phi$, so
we shall require the two functions $\varphi$ and $L$ to be first integrals of
some of the available $F$-generated vector fields. Explicitly, we require the
following:
\vskip 0.5cm
$1^0$. {\it The phase function}\
$\varphi$\  {\it is a first integral of the three vector fields}
${\bf A,A^*}$ and ${\bf R}$: ${\bf A}(\varphi)=0, {\bf A^*}(\varphi)=0,
{\bf R}(\varphi)=0$.
\vskip 0.5cm
$2^0$. {\it The scale factor} $L$ {\it is a
non-zero finite first integral of the vector field} ${\bf S}$: ${\bf
S}(L)=0$.
\vskip 0.5cm
The requirement ${\bf R}(\varphi)=0$ just means that
in these coordinates $\varphi$ does not depend on the coordinate $z$. The two
other equations of $1^0$ define the following system of differential
equations for $\varphi$:
\[
-\varphi \frac{\partial \varphi}{\partial
x}-\sqrt{1-\varphi^2}\frac{\partial \varphi}{\partial y}=0,\
\sqrt{1-\varphi^2}\frac{\partial \varphi}{\partial
x}-\varphi\frac{\partial \varphi}{\partial y}=0.
\]
Noticing that the matrix
\[
\left\|\matrix{
-\varphi              &-\sqrt{1-\varphi^2}\cr
\sqrt{1-\varphi^2}    &-\varphi
\cr}\right\|
\]
has non-zero determinant, we conclude that the only solution of the above
system is the zero-solution:
\[
\frac{\partial \varphi}{\partial x}=\frac{\partial \varphi}{\partial y}=0.
\]
We obtain that in the coordinates used the phase function $\varphi$ depends
only on $\xi$. Therefore, in view of (4), for the scale factor $L$ we get
$$
L=\frac{\sqrt{1-\varphi^2}}{|\varphi_\xi|}.
$$
Now, the requirement $2^0$, which in these coordinates reads
\[
{\bf S}(L)=\frac{\partial L}{\partial \xi}=
\frac{\partial}{\partial \xi}\frac{\sqrt{1-\varphi^2}}{|\varphi_\xi|}=0,
\]
just means that the scale factor $L$ is a pure constant: $L=const$. In this
way we obtain the differential equation
\begin{equation}
\frac{\partial \varphi}{\partial \xi}=
\mp \frac 1L \sqrt{1-\varphi^2}.
\end{equation}
The obvious solution to this equation reads
\begin{equation}
\varphi(\xi)=cos\left(\kappa\frac{\xi}{L}+const\right),
\end{equation}
where $\kappa=\pm 1$.
It worths to note that the naturally arising {\it characteristic frequency}
according to the equation
\begin{equation}
\nu=\frac cL,
\end{equation}
has nothing to do with the concept of frequency in CED. In fact, {\it the
quantity $L$ can not be defined in Maxwell's theory}.

Finally (recalling [1]) we note, that the so obtained phase function
$\varphi(\xi)$ leads to the following.
The 2-form $tr({\cal R}^0)$, where ${\cal R}^0$ is the matrix of
2-forms, formed similarly to the matrix ${\cal R}$, but using the basis
$({\bf A,\varepsilon A^*,R,S})$ instead of the basis
$(A,\varepsilon A^*,{\bf R,S})$, is closed. In fact,
\[
tr({\cal R}^0)=\varphi dx\wedge d\xi +\varphi dy\wedge d\xi -dy\wedge dz +
dz\wedge d\xi
\]
and since $\varphi=\varphi(\xi)$, we get ${\bf d}tr({\cal R}^0)=0$.
Note also that the above explicit form of $tr({\cal R}^0)$ allows to define
the phase function by
\[
\varphi=\sqrt{\frac{|tr({\cal R}^0)|^2}{2}}.
\]
This class of solutions we call {\it almost photon-like}.

\noindent{\it Remark}.\ If one of the two functions $u$ and $p$, for example
$p$, is equal to zero: $p=0$, then, formally, we again have a solution, which
may be called {\it linearly polarized} by obvious reasons. Clearly, the phase
function of such solutions will be {\it constant}: $\varphi=const$, so, the
corresponding scale factor becomes infinitely large: $L\rightarrow \infty$,
therefore, condition $2^0$ is not satisfied. The reason for this is, that at
$p=0$ the function $u$ becomes a {\it running wave} and we get \linebreak
$|\delta F|=|\delta *F|=0$, so the scale factor can not be defined by the
relation $L=|A|/|\delta F|$.

\section{Intrinsic angular momentum (helicity) \\ and
photon-like solutions}
The problem for describing the {\it intrinsic angular momentum} ($IAM$), or in
short {\it helicity, spin} of the photon is of fundamental importance in
modern physics, therefore, we shall pay a special attention to it. In
particular, we are going to consider two approaches for its mathematical
description. But first, some preceding comments.

First of all, {\it there is no any doubt that every free photon carries such
an  intrinsic angular momentum}. Since the angular momentum is a
conserved quantity, the existence of the photon's intrinsic angular
momentum can be easily established and, in fact, its presence  has been
experimentally proved by an immediate observation of its mechanical action
and its value has been numerically measured.  Assuming this is so, we have to
understand its origin, nature and its entire meaning for the existence and
outer relations of those natural entities, called shortly photons somewhere
in the first quarter of this century.

So, we begin with the assumption:{\it every free photon carries an intrinsic
angular momentum with integral value equal to the Planck's constant
$h$}. According to our understanding, the photon's IAM comes from an intrinsic
{\it periodic process}. This point of view undoubtedly leads to the notion,
that photons {\it are not} point-like structureless objects, they have a
structure, i.e. they are {\it extended objects}. In fact, according to one
of the basic principles of physics {\it all free objects move as a whole
uniformly}.  So, if the photon is a point-like object any characteristic of a
periodic process, e.g. frequency,  should come from an outside force field,
i.e. it can not be free: a free point-like (structureless) object can not
have the characteristic frequency.

This simple, but true, conclusion sets the theoretical physics of the first
quarter of this century faced with a serious dilemma: to keep the notion of
structurelessness and to associate in a formal way the characteristic
frequency to the microobjects, or to leave off the notion of structuelesness,
to assume the notion of extendedness and availability of intrinsically
occurring periodic process and to build corresponding integral
characteristics, determined by this periodic process. A look back in time
shows that the majority of those days physicists had adopted the first
approach, which has brought up to life quantum mechanics as a computing
method , and the dualistic-probabilistic interpretation as a philosophical
conception.  If we set aside the widespread and intrinsically controversial
idea that all microobjects are at the same time (point-like) particles and
(infinite) waves, and look impartially, in a fair-minded way, at the quantum
mechanical wave function for a {\it free} particle, we see that {\it the only
positive consequence} of its introduction is {\it the legalization of
frequency}, as an inherent characteristic of the microobject. In fact, the
probabilistic interpretation of the quantum mechanical wave function for a
free object, obtained as a solution of the free Schroedinger equation, is
impossible since its square is not an integrable quantity (the integral is
infinite). The frequency is really needed not because of the
dualistic-probabilistic nature of microobjects, it is needed because the
Planck's relation $E=h\nu$ turns out to be universally true in microphysics,
so there is no way to avoid the introduction of frequency. The question is,
if the introduction of frequency necessarily requires some (linear) "wave
equation" and the simple complex exponentials of the kind
$const.exp[i({\bf k.r}-\nu t)]$, i.e. running waves,
as "free solutions".  Our answer to this question is "no".  The
classical wave is something much richer and a much more engaging concept,
so it hardly worths to use it just because of the attribute of frequency.
In our opinion, it suffices to have a periodic process at hand.

These considerations made us turn to the soliton-like objects, they present
the two features of the microobjects ({\it localized spatial extendedness and
time-periodicity}), simultaneously, and, therefore, seem to be more adequate
theoretical models for those microobjects, obeying the Planck's relation
$E=h\nu$. Of course, if we are interested only in the behaviour of the
microobject as a whole, we can use the point-like notion, but any attempt to
give a meaning of its integral characteristics without looking for their
origin in the consistent intrinsic dynamics and structure, in our opinion, is
not a perspective theoretical idea. And the "stumbling point" of such an
approach is just the availability of an intrinsic mechanical angular
momentum, which can not be understood as an attribute of a free structureless
object.

\vskip 0.5cm
Having in view the above considerations, we are going to consider two ways to
introduce and define the intrinsic angular momentum as a local quantity and
to obtain, by integration, its integral value. So, these two approaches will
be of use only for the spatially finite nonlinear solutions of our equations.
The both approaches introduce in different ways 3-tensors (2-covariant and
1-contravariant). Although these two 3-tensors are built of quantities,
connected in a definite way with the field $F$, their nature is quite
different. The first approach is based on an appropriate tensor
generalization of the classical Poynting vector. The second approach makes
use of the concept of {\it torsion}, connected with the field $F$, considered
as 1-covariant and 1-contravariant tensor. The first approach is pure
algebraic, while the second one uses derivatives of $F_{\mu\nu}$. The
 spatially finite nature of the solutions $F$ allows to
build corresponding integral conserved quantities, naturally interpreted as
angular momentum. The scale factor $L$ appears as a multiple, so these
quantities go to infinity for all linear (i.e. for Maxwell's) solutions.

In the first approach we make use of the scale factor $L$, the isotropic
vector field ${\bf V}$ and the two 1-forms $A$ and $A^*$. By these four
quantities we build the following 3-tensor $H$:
\begin{equation}
H=\kappa \frac Lc {\bf V}\otimes(A\wedge A^*).
\end{equation}
The connection with the classical vector of Poynting comes through the
exteriour product of $A$ and $A^*$, the 3-dimensional sense of which is just
the Pointing's vector. In components we have
\[
H^\mu_{\nu\sigma}= \kappa\frac Lc {\bf V}^\mu(A_\nu A^*_\sigma-A_\sigma
A^*_\nu).
\]
In our system of coordinates we get
\[
H=\kappa\frac Lc\left(-\varepsilon\frac{\partial}{\partial z}+
\frac{\partial}{\partial \xi}\right)\otimes(\varepsilon\phi^2dx\wedge dy),
\]
so, the only non-zero components are
\[
H^3_{12}=-H^3_{21}=-\kappa\frac Lc \phi^2,
\ H^4_{12}=-H^4_{21}=\kappa\varepsilon\frac Lc \phi^2.
\]
It is easily seen, that the divergence
$\nabla_\mu H^\mu_{\nu\sigma}\rightarrow\nabla_\mu H^\mu_{12}$
is equal to 0. In fact,
\[
\nabla_\mu H^\mu_{12}=\frac{\partial}{\partial z}H^3_{12}+
\frac{\partial}{\partial \xi}H^4_{12}=\kappa\frac Lc\left[-(\phi^2)_z+
(\varepsilon \phi^2)_\xi\right]=0
\]
because $\phi^2$ is a running wave along the coordinate $z$. Since the
tangent bundle is trivial we may construct the antisymmetric 2-tensor
\[
{\bf H}_{\nu\sigma}=\int_{R^3}{H_{4,\nu\sigma}}dxdydz,
\]
the constant components of which are conserved quantities.
\[
{\bf H}_{12}=-{\bf H}_{21}=\int_{R^3}{H_{4,12}}dxdydz
=\kappa\varepsilon\frac Lc W=\kappa\varepsilon WT=
\kappa\varepsilon \frac W\nu.
\]
The non-zero eigen values of  ${\bf H}_{\nu\sigma}$ are pure imaginary and
are equal to $\pm iWT$. This tensor has unique non-zero invariant $P(F)$,
\begin{equation}
P(F)=\sqrt{\frac12 {\bf H}_{\nu\sigma}{\bf H}^{\nu\sigma}}=WT.
\end{equation}
The quantity $P(F)$ will be called {\it Planck's invariant} for the finite
nonlinear solution $F$. All finite nonlinear solutions $F_1,F_2,,...$,
satisfying the condition
\[
P(F_1)=P(F_2)=...=h,
\]
where $h$ is the Planck's constant, will be called further {\it photon-like}.
The tensor field $H$ will be called {\it intrinsic angular momentum tensor}
and the tensor ${\bf H}$ will be called {\it spin tensor} or {\it helicity
tensor}. The Planck's invariant $P(F)=WT$, having the physical dimension of
action, will be called {\it integral angular momentum}, or just {\it spin} or
{\it helicity}.

The reasons to use this terminology are quite clear: the time evolution of
the two mutually orthogonal vector fields $A$ and $A^*$ is a
rotational-advancing motion around and along the $z$-coordinate (admissible
are the right and the left rotations: $\kappa =\pm 1$) with the advancing
velocity of $c$ and the frequency of circulation $\nu=c/L$. We see the basic
role of the two features of the solutions: their soliton-like character,
giving finite value of all integral quantities, and their nonlinear
character, allowing to define the scale factor $L$ correctly. From this point
of view the intrinsic angular momentum $h$ of a free photon is far from being
{\it incomprihensible} quantity, connected with the even more
incomprihensible duality "wave-particle", and it looks as a quite normal
integral characteristic of a solution, presenting a model of our knowledge of
the free photon.

\vskip 0.5cm
We proceed to the second approach to introduce $IAM$ by recalling the
definition of {\it torsion} of two (1,1) tensors. If $G$ and $K$ are 2 such
tensors

\[
G=G_\mu
^\nu dx^\mu\otimes \frac{ \partial}{\partial x^\nu },\quad K=K_\mu ^\nu
dx^\mu\otimes \frac{ \partial}{\partial x^\nu },
\]
their torsion is defined
as a 3-tensor $S_{\mu\nu}^\sigma=- S_{\nu\mu}^\sigma $ by the equation
\[
S(G,K)(X,Y)=[GX,KY]+[KX,GY]+GK[X,Y]+KG[X,Y]-
\]
\[
-G[X,KY]-G[KX,Y]-K[X,GY]-K[GX,Y],
\]
where $[,]$ is the Lie-bracket of vector fields,
\[
GX=G_\mu ^\nu X^\mu \frac {\partial}{\partial x^\nu},\quad
GK=G_\mu ^\nu K_\sigma ^\mu dx^\sigma \otimes \frac{\partial}{\partial x^\nu}
\]
and $X,Y$ are 2 arbitrary vector fields. If $G=K$, in general
 $S(G,G)\neq 0$ and
\[
S(G,G)(X,Y)=2\left\{[GX,GY]+GG[X,Y]-G[X,GY]-G[GX,Y]\right\}.
\]
This last expression defines at every point $x\in M$ the torsion $S(G,G)=S_G$
of $G$ with respect to the 2-dimensional plain, defined by the two vectors
$X(x)$ and $Y(x)$. Now we are going to compute the torsion $S_F$ of the
nonlinear solution $F$ with respect to the intrinsically defined by the two
unit vectors ${\bf A}$ and ${\bf \varepsilon A^*}$ 2-plain. In components we
have
\[
(S_F)_{\mu \nu }^\sigma =2\left[ F_\mu ^\alpha \frac{\partial F_\nu
^\sigma} {\partial x^\alpha }-F_\nu ^\alpha \frac{\partial F_\mu ^\sigma
}{\partial x^\alpha }-F_\alpha ^\sigma \frac{\partial F_\nu ^\alpha
}{\partial x^\mu } +F_\alpha ^\sigma \frac{\partial F_\mu ^\alpha }{\partial
x^\nu }\right].
\]
In our coordinate system
\[
{\bf A}=-\varphi
\frac{\partial} {\partial x}-\sqrt{1-\varphi ^2} \frac{\partial}{\partial
y},\quad {\bf \varepsilon A^*} =
\sqrt{1-\varphi ^2}\frac{\partial}{\partial x}-
\varphi \frac{\partial} {\partial y},
\]
so,
\[
(S_F)_{\mu \nu }^\sigma{\bf A}^\mu {\bf \varepsilon A^*}^\nu =
(S_F)_{12}^\sigma ({\bf A}^1{\bf \varepsilon A^*}^2-
{\bf A}^2{\bf \varepsilon A^*}^1).
\]
For $(S_F)_{12}^\sigma $ we get
\[
(S_F)_{12}^1=(S_F)_{12}^2=0,\quad
(S_F)_{12}^3=-\varepsilon (S_F)_{12}^4=2\varepsilon \{p(u_\xi -\varepsilon
u_z)-u(p_\xi -\varepsilon p_z)\}.
\]
\vskip 0.5cm
{\it Remark}. In our case
$(S_F)_{12}^\sigma=(S_{*F})_{12}^\sigma $, so further we shall work with
$S_F$ only.
\vskip 0.5cm
It is easily seen that the following relation holds:
${\bf A}^1{\bf \varepsilon A^*}^2-{\bf A}^2{\bf \varepsilon A^*}^1=1.$
Now, for the almost photon-like solutions
\[
u=\phi (x,y,\xi +\varepsilon z)\cos\left(\kappa \frac \xi L +const\right),
\quad
p=\phi (x,y,\xi +\varepsilon z)\sin \left(\kappa \frac \xi L +const\right)
\]
we obtain
\[
(S_F)_{12}^3=-\varepsilon (S_F)_{12}^4=
-2\varepsilon \frac \kappa L \phi ^2,
\]

\[
(S_F)_{\mu \nu }^\sigma {\bf A}^\mu {\bf \varepsilon A^*}^\nu
=\left[0,0,-2\varepsilon
\frac \kappa L \phi ^2,2\frac \kappa L \phi ^2\right].
\]
Since $\phi^2$ is a running wave along the $z$-coordinate, the vector
field $S_F({\bf A,\varepsilon A^*})$ has zero divergence:
$\nabla_\nu \left[S_F({\bf A,\varepsilon A^*})\right]^\nu=0$.
Now we define the {\it helicity vector} for the solution $F$ by
\[
\Sigma_F=\frac{L^2}{2c}S_F({\bf A,\varepsilon A^*}).
\]
Since $L=const$, then $\Sigma_F$ has also zero divergence, so the integral
quantity
\[
\int{\left(\Sigma_F\right)_4}dxdydz
\]
does not depend on time and is equal to $\kappa WT$. The photon-like
solutions are separated in the same way by the condition $WT=h$. Here are
three more integral expressions for the quantity $WT$. We form the 4-form
\[
-\frac 1L {\bf S}\wedge*\Sigma_F=\frac \kappa c \phi^2\omega_\circ
\]
and integrate it over the 4-volume ${\cal R}^3\times L$, the result is
$\kappa WT$. Besides, we verify easily the relations
\[
\frac 1c \int_{R^3\times L}|A\wedge
A^*|\omega_\circ=
\frac{L^2}{c}\int_{R^3\times L}|\delta F\wedge \delta *F|\omega_\circ=WT.
\]

Since we separate the photon-like solutions by the relation $WT=h$, the last
expressions suggest the following interpretation of the Planck's constant
$h$.  Since $|A\wedge A^*|$ is proportional to the area of the square,
defined by the two mutually orthogonal vectors $A$ and $\varepsilon A^*$, the
above integral sums up all these areas over the whole 4-volume, occupied by
the solution $F$ during the intrinsically determined time period $T$, in
which the couple $(A,\varepsilon A^*)$ completes a full rotation. The same
can be said for the couple $(\delta F,\delta *F)$ with some different factor
in front of the integral. This shows quite clearly the "helical" origin of
the full energy $W=h\nu$ of the single photon.

\section{Solutions in spherical cordinates}

The so far obtained soliton-like
solutions describe objects, "coming from infinity" and
"going to infinity". Of interest are also soliton like solutions "radiated"
from, or "absorbed", by some central "source" and propagating radially from
or to the center of this source. We are going to show, that our equations [1]
admit such solutions too. We assume this central source to be a small ball
$R^0$ with radius $r_\circ$, and put the origin of the coordinate system at
the center of the source-ball. The standard spherical coordinates
$(r,\theta,\varphi,\xi)$ will be used and all considerations will be carried
out in the region out of the ball $R^0$. In these coordinates we have
\[
ds^2=-dr^2-r^2d\theta ^2-r^2 sin^2\theta d\varphi^2 +d\xi^2, \
\sqrt{|\eta|}=r^2sin\theta.
\]
The $*$-operator acts in these coordinates as follows:
\[
\begin{array}{ll}
*dr=r^2 sin\theta d\theta\wedge d\varphi\wedge d\xi  &*(dr\wedge
d\theta\wedge d\varphi )=(r^2 sin\theta)^{-1} d\xi \cr
*d\theta=-sin\theta dr\wedge d\varphi\wedge d\xi     &*(dr\wedge
d\theta\wedge d\xi)=sin\theta d\varphi  \cr
*d\varphi=(sin\theta)^{-1}dr\wedge d\theta d\xi      &*(dr\wedge
d\varphi\wedge d\xi)=-(sin\theta)^{-1}d\theta     \cr
*d\xi=r^2 sin\theta dr\wedge d\theta d\varphi        &*(d\theta\wedge
d\varphi\wedge d\xi)=(r^2 sin\theta)^{-1} dr
\end{array}
\]
\[
\begin{array}{ll}
*(dr\wedge d\theta)=-sin\theta d\varphi\wedge d\xi        &*(d\theta\wedge
d\varphi)=-(r^2 sin\theta)^{-1}dr\wedge d\xi \cr
*(dr\wedge d\varphi)=(sin\theta)^{-1} d\theta\wedge d\xi  &*(d\theta\wedge
d\xi)=-sin\theta dr\wedge d\varphi \cr
*(dr\wedge d\xi)=r^2 sin\theta d\theta\wedge d\varphi
&*(d\varphi\wedge d\xi=(sin\theta)^{-1}dr\wedge d\theta.
\end{array}
\]
We look for solutions of the following kind:
\begin{equation}
F=\varepsilon udr\wedge d\theta +ud\theta\wedge d\xi +\varepsilon
pdr\wedge d\varphi
+pd\varphi \wedge d\xi,
\end{equation}
where $u$ and $p$ are spatially finite functions. We get
\[
*F=\frac{p}{sin\theta}dr\wedge d\theta +\varepsilon \frac{p}{sin\theta}
d\theta\wedge d\xi-
usin\theta dr\wedge d\varphi-\varepsilon sin\theta d\varphi \wedge d\xi.
\]
The following relations hold:
\[
F\wedge F=2\varepsilon(up-up)dr\wedge d\theta\wedge d\varphi \wedge d\xi=0,
\]
\[
F\wedge *F= \left(-u^2 sin\theta +u^2 sin\theta-
\frac{p^2}{sin\theta}+\frac{p^2}{sin\theta}\right)dr\wedge d\theta\wedge
d\varphi\wedge d\xi=0,
\]
i.e. the two invariants are equal to zero: $(*F)_{\mu\nu}F^{\mu\nu}=0,
\ F_{\mu\nu}F^{\mu\nu}=0$.

After some elementary computation we obtain
\[
\delta F\wedge F=\delta *F\wedge *F=
\varepsilon\left[u\left(\varepsilon p_r +p_\xi\right)-
p\left(\varepsilon u_r +u_\xi\right)\right]dr\wedge d\theta\wedge d\varphi+
\]
\[
+\left[u\left(\varepsilon u_r +u_\xi\right)-
u\left(\varepsilon p_r +p_\xi \right)\right]d\theta\wedge d\varphi\wedge d\xi,
\]
\[
F\wedge *{\bf d}F=\varepsilon\left[u\left(\varepsilon u_r
+u_\xi\right)sin\theta+
\frac{p\left(\varepsilon p_r +p_\xi \right)}{sin\theta}\right]dr\wedge
d\theta\wedge d\varphi-
\]
\[
-\varepsilon\left[u\left(\varepsilon u_r + u_\xi\right)sin\theta+
\frac{p\left(\varepsilon p_r + p_\xi
\right)}{sin\theta}\right]d\theta\wedge d\phi\wedge d\xi,
\]
\[
(*F)\wedge *{\bf d}*F=\left[u\left(\varepsilon u_r +u_\xi\right)sin\theta+
\frac{p\left(\varepsilon p_r +p_\xi \right)}{sin\theta}\right]dr\wedge
d\theta\wedge d\varphi-
\]
\[
-\left[u\left(\varepsilon u_r + u_\xi\right)sin\theta+
\frac{p\left(\varepsilon p_r + p_\xi \right)}{sin\theta}\right]d\theta\wedge
d\phi\wedge d\xi.
\]
So, the two functions $u$ and $p$ have to satisfy the equation
\begin{equation}
u\left(\varepsilon u_r + u_\xi\right)sin\theta+
\frac{p\left(\varepsilon p_r + p_\xi \right)}{sin\theta}=0,
\end{equation}
which is equivalent to the equation
\begin{equation}
\left(u^2 sin\theta +\frac{p^2}{sin\theta}\right)_\xi +
\varepsilon\left(u^2 sin\theta +\frac{p^2}{sin\theta}\right)_r=0.
\end{equation}
The general solution of this equation is
\begin{equation}
u^2 sin\theta +\frac{p^2}{sin\theta}=\phi^2(\xi - \varepsilon r,\theta,\phi).
\end{equation}
For the non-zero components of the energy-momentum tensor we obtain
\begin{equation}
-Q_{1}^{1}=-Q_{1}^{4}=Q_{4}^{1}=Q_4^4=
\frac{1}{4\pi r^2 sin\theta}\left(u^2 sin\theta +\frac{p^2}{sin\theta}\right).
\end{equation}
It is seen that the energy density is not exactly a running wave but when we
integrate to get the integral energy, the integrand is exactly a running
wave:
\[
W=\frac{1}{4\pi}\int_{R^3-R^0}{*\left(Q_\mu^4 d\xi\right)}=
\frac{1}{4\pi}\int_{R^3-R^0}{\left(u^2 sin\theta
+\frac{p^2}{sin\theta}\right)}dr\wedge d\theta\wedge d\phi.
\]
Since the functions $u$ and $p$ are spatially finite, the integral energy $W$
is finite, and from the explicit form of the energy-momentum tensor it
follows the well known relation between the integral energy and momentum:
$\ W^2-c^{2} {\bf p}^2=0$.

\newpage
{\bf References}:\\

1. Donev,S., Tashkova,M., {\it \bf Extended Electrodynamics}: II.
{\it Properties and Invariant Characteristics of the Non-linear Vacuum
Solutions}, submitted for publication.

 \end{document}